\documentclass[12pt]{article}
\usepackage{amsmath}
\usepackage{array}
\usepackage{amsfonts}
\newcommand{\ga}{\alpha}
\newcommand{\gb}{\beta}
\renewcommand{\gg}{\gamma}
\newcommand{\gd}{\delta}

\newcommand{\gx}{\xi}

\newcommand{\gr}{\rho}

\newcommand{\gs}{\sigma}

\newcommand{\go}{\omega}

\newcommand{\gp}{\pi}
\newcommand{\gps}{\psi}
\newcommand{\get}{\eta}

\newcommand{\Bga}{{\boldsymbol \alpha}}

\newcommand{\Bge}{{\boldsymbol \epsilon}}

\newcommand{\Bgt}{{\boldsymbol \tau}}

\newcommand{\gF}{\Phi}

\newcommand{\cM}{{\cal M}}

\newcommand{\uga}{{\underline \alpha}}

\newcommand{\bs}{{\bar s}}

\newcommand{\bv}{{\bar v}}

\newcommand{\bz}{{\bar z}}

\newcommand{\bX}{{\bar X}}

\newcommand{\tr}{\text{tr}}

\newcommand{\Id}{\text{\small 1}\hspace{-3.5pt}\text{1}}

\newcommand{\lra}{\longrightarrow}

\newcommand{\der}{\partial}

\newcommand{\inv}{^{-1}}

\newcommand{\nit}{\noindent}
\newcommand{\nl}{\newline}
\newcommand{\np}{\newpage}

\newcommand{\scp}{\scriptstyle}

\newcommand{\undr}[1]{{\underline{#1}}}

\newcommand{\labl}[1]{\label{#1}}
\newcommand{\ESO}{E_6/SO(10)\times U(1)}

\newcommand{\Kh}{K\"{a}hler}

\newcommand{\Intr}{\mathbb{Z}}
\newcommand{\Cplx}{\mathbb{C}}
\newcommand{\Real}{\mathbb{R}}
\newcommand{\beq}{\begin{equation}}
\newcommand{\eeq}{\end{equation}}
\newcommand{\barr}{\begin{array}}
\newcommand{\earr}{\end{array}}
\newcommand{\equ}[1]{\begin{gather} #1 \end{gather}}

\newcommand{\mtrx}[1]{\begin{matrix} #1 \end{matrix}}
\newcommand{\pmtrx}[1]{\begin{pmatrix} #1 \end{pmatrix}}

\newcounter{oldcounter}

\newcommand{\be}{\begin{equation}}
\newcommand{\ee}{\end{equation}}
\newcommand{\bea}{\begin{eqnarray}}
\newcommand{\eea}{\end{eqnarray}}

\begin{document}

\pagestyle{empty}
\markright{NIKHEF/99-025}
\begin{flushright}
NIKHEF/99-025 \\
\end{flushright}
\vspace{4ex}
\begin{center}
{\Large {\bf Line bundles in supersymmetric coset models}} \\
\vspace{5ex}

{\large S.\ Groot Nibbelink}\\
\vspace{3ex}

{\large NIKHEF, P.O.\ Box 41882}\\
\vspace{3ex}

{\large 1009 DB, Amsterdam, The Netherlands} \\
\vspace{5ex}
{\em Revised version}\\
\end{center}
\vspace{10ex}
%
%
\nit
{\small 
{\bf Abstract}\nl
The scalars of an $N = 1$ supersymmetric $\sigma$-model in $4$ 
dimensions parameterize a K\"ahler manifold. 
The transformations of their fermionic superpartners 
under the isometries are often anomalous. 
These anomalies can be canceled by introducing additional chiral 
multiplets with appropriate charges. To obtain the right charges 
a non-trivial singlet compensating multiplet can be used. 
However when the topology of the underlying \Kh\ manifold is 
non-trivial, the consistency of this multiplet 
requires that its charge is quantized. 
This singlet can be interpreted as a section of a line bundle. 
We determine the K\"ahler potentials corresponding to the
minimal non-trivial
singlet chiral superfields for any compact K\"ahlerian coset space $G/H$.
The quantization condition may be in conflict with the requirement of 
anomaly cancelation. 
To illustrate this, we discuss the consistency  of anomaly free 
models based on the coset spaces 
$E_6/SO(10) \times U(1)$ and 
$SU(5)/SU(2) \times U(1) \times SU(3)$.
}
\vspace{10ex}
{\small
}
\np

\pagestyle{plain}
\pagenumbering{arabic}

%
%
Non-linear $\gs$-models in 4 dimensions with $N=1$ supersymmetry 
may be the effective field theory setting for the discussion 
of models beyond the standard model.
Chiral fermions $\gps_L^\ga$ are contained in an irreducible 
representation of $N=1$ supersymmetry called chiral multiplets 
$\gF^\ga = (z^\ga, \gps_L^\ga, h^\ga)$. 
The lowest components of those multiplets $z^\ga$ are complex 
scalars that are the coordinates of a complex \Kh\ manifold 
$\cM$ \cite{Bagger:1982}.  
Locally the metric $g_{\uga\ga}$ of this manifold is obtained 
from a \Kh\ potential $K(z, \bz)$. 
Non-trivial examples of \Kh\ manifolds are provided by 
homogeneous coset spaces $G/H$, where $G$ is any compact Lie 
group and $H$ is the centralizer of a torus in $G$. 
\par
The chiral fermion content of a supersymmetric model based on 
the \Kh\ manifold $\cM$ is often anomalous 
\cite{Moore:1984dc,Aoyama:1985er}. 
This leads to the conclusion that supersymmetric models based 
on $G/H$ cosets are inconsistent unless a complete mirror sector 
is introduced, which is phenomenological uninteresting as it 
easily results in large fermion masses 
\cite{Buchmuller:1987zp,Kotcheff:1990ck}.
However one can avoid this by including additional non-mirror 
(matter) chiral multiplets to cancel anomalies. 
In order that these matter multiplets respect all isometries 
of $\cM$ they can be introduced as tensor products of 
co(ntra)variant vectors on $\cM$ 
\cite{vanHolten:1985xk,Achiman:1984ku}. 
This type of matter alone is often not sufficient to build 
anomaly free models: the charge assignment is to restrictive. 
To overcome this difficulty, a singlet chiral multiplet 
 was introduced in ref.\ \cite{GrootNibbelink:1998tz}, that 
transforms non-trivially under the isometries of $\cM$.
This non-trivial singlet superfield can be used as a compensator 
to give physical chiral multiplets the charges required for 
isometry anomaly cancelation. 
For this non-trivial singlet multiplet, as well as for the other 
matter multiplets, invariant \Kh\ potentials have to be constructed.
From these one obtains a supersymmetric lagrangean, that 
is invariant under the action of the isometries of $\cM$. 
\par
The definitions of matter chiral superfields, as given above, 
take only the local properties of the \Kh\ manifold $\cM$ into 
account. 
In order to guarantee that the definitions of these matter multiplets 
do not lead to global inconsistencies, additional 
constraints have to be imposed.
In mathematics a well-defined function on a manifold is called section 
of a bundle over this manifold. 
\cite{Wells:80,Nakahara:1990th}. 
In this language, a globally well-defined matter multiplet has a  
scalar component that is a section of a bundle.
A co(ntra)variant vector on $\cM$ can be interpreted globally as 
a section of the (co)tangent bundle and is therefore well-defined.
The global definition of the non-trivial singlet compensating 
multiplet is more involved: when the topology of the \Kh\ manifold 
$\cM$ is non-trivial, there is an additional quantization 
condition \cite{Wells:80}, called the cocycle condition. 
The corresponding bundle is called a complex line bundle.
This quantization condition is equivalent to the requirement, that the 
integral over the \Kh\ form associated with the \Kh\ potential 
of the non-trivial singlet is equal to $2\pi \Intr$. 
In particular there is a minimal charge; all charges 
of non-trivial singlets have to be 
an integer number times the minimal charge. 
This condition leads to quantization of Newton's constant in 
supergravity theories \cite{Witten:1982hu}, when the \Kh\ 
potential is covariant but not invariant. 
The condition also restricts or can even be in conflict with 
the freedom of the charge 
assignment which was used in refs.\  
\cite{GrootNibbelink:1998tz,GrootNibbelink:1999dt} to obtain 
anomaly free supersymmetric models based on the coset $G/H$.
\par
%
%
The objectives of this letter are the following. 
We want to determine the minimal non-trivial singlet multiplets 
that can be coupled to compact \Kh ian coset spaces $G/H$.
Such a singlet multiplet can be used in supersymmetric model building 
to obtain the appropriate charge assignment for anomaly 
cancelation. 
To this end we have to identify the minimal line bundles, which 
can be done by showing that the corresponding \Kh\ form satisfies 
a minimal integral condition. 
To obtain general results, we introduce some Lie group 
machinery to describe $G/H$ coset spaces 
for any compact simple Lie group $G$. 
Using this we review the construction of \Kh\ potentials for 
these homogeneous spaces following refs.\  
\cite{Itoh:1986ha,Bando:1988br} with a few minor changes in 
notation. 
We obtain a generating set of \Kh\ potentials $K^J$ that satisfy 
the minimal cocycle condition and which may be used to 
construct invariant lagrangeans for non-trivial singlets superfields.
The last step is to determine the charges of these superfields. 
It is essential to know these charges as they may be such that
anomaly cancelation is not possible. 
To show that is a non-trivial requirement, we investigate whether 
the construction of two anomaly free models is compatible with the 
cocyle condition:  $\ESO$ \cite{GrootNibbelink:1998tz} and a 
Grassmannian version of the standard model 
\cite{GrootNibbelink:1999dt}. 
In the latter model matter is introduced by using that the 
transformation rules of covariant vectors factorizes. 
For completeness, we check the consistency of these matter 
representations as well. 
\\[3mm]
%
%
We start by discussing the quantization condition of the charge of 
non-trivial singlet representation when the topology of $\cM$ is 
non-trivial: the cocycle condition for a complex line bundle. 
By using the covariance of a \Kh\ potential $K$ 
\equ{
K(z', \bz') = K(z, \bz) - c(z) - c^\dag(\bz)
} 
under the isometries of $\cM$, the transformation rule 
\equ{
s' =  e^{-c(z)} s
\labl{LineBundle}
}
of the scalar component $s$ of a non-trivial chiral multiplet 
is obtained. The invariant \Kh\ potential  
\equ{
K_{\text{line}} (s, \bs; z, \bz) = 
\bs e^{-K(z, \bz)} s.
\labl{LinePot}
}
can be used for supersymmetric model building purposes
 \cite{GrootNibbelink:1998tz}.
\par
A well-defined complex line 
bundle over a manifold satisfies consistency 
conditions \cite{Wells:80} that insure the global existence of
this non-trivial singlet representation. 
In order that this singlet can be interpreted as a section 
of a line bundle, the cocycle condition 
on three overlapping coordinate patches $(1,2,3)$ of $\cM$ 
demands that
\equ{
c^{(123)} \equiv c^{(12)} + c^{(23)} + c^{(31)} =
 - 2 \gp i \Intr.
} 
Theorems by de Rham \cite{Wells:80,Bott:82} tell us that this 
requirement is equivalent to the condition that the integral 
\(
\frac 1 {2 \gp} \int_C \go(K) = \Intr
\)
of any \Kh\ $2$-form $\go(K)$ over any $2$-cycle $C$ of $\cM$ is 
integer. Locally the \Kh\ form $\go(K)$ can be obtained from a \Kh\ 
potential $K$ as 
\equ{
\go(K) = -i K_{,\uga\ga} d \bz^\uga \wedge d z^\ga.
\labl{KahlerForm}
}
In this letter we want to obtain the minimal charges of 
non-trivial singlet matter couplings to coset spaces $G/H$.
To do this we need to identify  a set of \Kh\ potentials $K^J$ 
and a set of $2$-cycles $C_I$ such that
\equ{
\int_{C_I} \go(K^J) = 2\gp \gd^J_I.
\labl{Integrability}
}
\\[3mm]
%
%
From now on we are only concerned with \Kh ian 
homogeneous coset 
spaces $G/H$ where $G$ a simple compact Lie group. 
It is convenient to use the Cartan 
normalization of the algebra of $G$ which we now review 
\cite{Cahn:1985wk,Baeuerle:1990sm,Broecker:85}.
$T = \{\Bgt_i\}$ is the set of generators of a Cartan subgroup of $G$. 
According to a theorem by Borel \cite{Borel:54} a coset $G/H$ 
is \Kh ian if $H$ is the centralizer $C(Y)$ of a torus in $G$.
This torus is generated by the set 
\equ{
Y = \{{\bf Y}^I = (G\inv)^{Ii} \Bgt_i\}.
}
(Hence the index $i$ enumerates the elements of $T$ and $I$ 
the elements of $Y$, hence the indices $I$ are a subset of $i$.)
An arbitrary linear combination of $\Bgt_i$ is denoted by $\Bgt$.
Let $\Bge_{\Bga}$ be the generator associated with root $\Bga$ 
and let $\Bge_{\pm i} = \Bge_{\pm \Bga_i}$ denote the creation 
and annihilation operators of the simple root $\Bga_i$. 
Hermitean conjugation of the generators is taken to give 
\(
\Bgt_i^\dag = \Bgt_i
\quad \text{and} \quad  
\Bge_\Bga^\dag = \Bge_{-\Bga}.
\) 
The algebra of $G$ associated with the simple roots 
can be stated as
\equ{
\mtrx{
[\Bgt_i, \Bgt_j] = 0, & 
[\Bge_i, \Bge_{-j}] = \gd_{ij} \Bgt_j, &
[\Bgt_i, \Bge_{\pm j}] = \pm G_{ij} \Bge_{\pm j},
}
\labl{CartanAlg}
}
where 
\(
G_{ij} \equiv 
{\scp \frac 
{ 2 \langle \Bga_i, \Bga_j \rangle }
{   \langle \Bga_i, \Bga_i \rangle } 
}
= \Bga_j ( \Bgt_i )
\)
is the Cartan matrix in this normalization.
%
%
%
Following \cite{Itoh:1986ha,Bando:1984cc,Bando:1984ab} we 
divided the generators of $G$ into the following sets. 
We have already defined the set $Y = \{{\bf Y}^I\}$.
We denote by 
\equ{
S = \{ S_a \} =  \{\Bgt_i | i \neq I \} \cup 
\{\Bge_\Bga | \Bga = \sum_{i \neq I} \ga^i \Bga_i \text{ a root} 
\}
}
the set of generators of $H$ that are not in $Y$. 
The two sets 
\equ{
X = \{ X_\Bga \} =
\{ \Bge_\Bga | \Bge_\Bga \notin S, \Bga > 0 \}
\quad \text{and} \quad
\bX = \{ \bX_\Bga \} =
\{ \Bge_\Bga | \Bge_\Bga \notin S, \Bga < 0 \}
}
contain the remaining part of the generators of $G$.
\par
%
%
Using this notation we give representations of elements of 
$H$, $G/H$, etc.
An element $h \in H$ is written as
\(
h = e^{i \gb S} e^{i\gg Y},
\)
where $\gb^a, \gg_I \in \Real$ and summation over indices has 
been assumed to be understood.
The subset $\hat H \subset G^\Cplx$ of the complexification of 
the group $G$ is generated by $Y, S$ and $X$. 
Any element $\hat h \in \hat H$ of $\hat H$ is represented as
\(
\hat h = e^{aX} e^{b S} e^{c Y},
\)
where $a^\Bga,b^a,c_I \in \Cplx$ and therefore an element of 
$G^\Cplx / \hat H$ can be written as 
\equ{
\gx (z) = e^{z \bX},
}
where $z^{\Bga} \in \Cplx$ carries the same root-indices as 
$\bX_\Bga$. 
And finally let $U(z, \bz)$ denote an element of the unitary 
representation of $G/H$. 
According to \cite{Itoh:1986ha} $\hat H$ is chosen such that 
$G/H \cong G^\Cplx/\hat H$, hence $U(z, \bz)$ can be expressed 
in terms of $\gx(z)$ as
\equ{
U(z, \bz) = \gx(z) 
e^{A(z, \bz) X} e^{B(z,\bz) S} e^{- \frac 12 K(z, \bz) Y}.
\labl{Uxi}
}
The representative $U(z,\bz)$ of the equivalence classes of 
$G/H$ is chosen such that $B(z, \bz)$ an $K_J(z, \bz)$ are real 
functions.
(The normalization of the functions $K(z, \bz)$ is chosen for 
convenience later on.)
\par
%
%
The non-linear transformation properties of the coordinates $z$ 
and $\bz$ of $G/H$ can be defined using $\gx$ 
\cite{Bando:1984ab,Bando:1984cc} or $U$ 
\cite{Coleman:69,Callan:69} by
\equ{
g \gx (z) = \gx(z') \hat h (z; g) 
\quad \text{and} \quad
g U(z, \bz) = U(z', \bz') h(z, \bz; g)
\labl{TransUxi}
}
for any element $g$ of $G$.
The functions $\hat h$ and $h$ are chosen such that $\gx(z')$ 
and $U(z', \bz')$ are again of the forms given above.
Combining these transformation rules with the identification 
of $G/H$ with $G^\Cplx/\hat H$ according to eq.\ \eqref{Uxi} 
shows that
\equ{
K_J(z', \bz') = K_J(z, \bz)  
- c_J(z;g) - {c_J}^\dag (\bz; g);
\labl{KJtrans}
}
thus the functions $K_J(z, \bz)$ transform as \Kh\ potentials 
\cite{Itoh:1986ha}.
There it is shown also that the set of \Kh\ 
potentials $\{ K_J \}$ is complete.
\\[3mm]
%
%
First of all we want to obtain an explicit formula for a fundamental 
\Kh\ potential $K_J$. 
To this end we consider a representation with orthonormal weight vectors 
$|{{\bf b}^J, {\bf w}}\rangle$ and highest weight 
${\bf b}^J$, that has all its Dynkin labels zero except for the $J$th 
one which is 1: 
\(
( b^J )_j = \gd^J_j.
\)
The highest weight vector $|{{\bf b}^J, {\bf b}^J}\rangle$ forms a 
one dimensional irreducible $H$-representation, 
as only generator that acts non-trivially on it ($\Bge_{-J}$), 
is not contained in $Y$ or $S$. 
Using the BKMU-projector
\(
P^J \equiv 
|{{\bf b}^J, {\bf b}^J}\rangle \langle {{\bf b}^J, {\bf b}^J}|,
\labl{PJproj}
\) 
that projects on this one-dimensional subspace, 
the fundamental \Kh\ potentials $K^J$ can be represented as 
\equ{
K^J(z, \bz) =K_I (G\inv)^{IJ} =  \ln {\det}_{\, P^J} 
\left[ \gx^\dag (\bz) \gx(z) \right].
\labl{KJ}
}
This follows because any \Kh\ potential 
\(
K_{\get_{\bf b}^{\;}}(z, \bz) 
\)
obtained using 
a BKMU-projector $\get_{\bf b}$ acting a representation with 
highest weight ${\bf b}$, can be decomposed 
\cite{Itoh:1986ha,Bando:1988br,Bando:1984ab} as
\equ{
K_{\get_{\bf b}^{\;}}(z, \bz) 
\equiv
\ln {\det}_{\,\get_{\bf b}^{\;}}
\left[
\gr_{\bf b}^{\:}( \gx^\dag( \bz ) )
\gr_{\bf b}^{\:}( \gx( z ) )
\right]
= \tr_{\bf b}^{\;} \Bigl( \get_{\bf b}^{\;} \Bgt_I^{\;} \Bigr)
K^I(z, \bz) .
\labl{ProjKahlerKI}
}
For the BKMU-projection operators $P^J$ we find that
\(
\tr \left( P^J \Bgt_I^{\;} \right) = \gd^J_I,
\labl{PJcharge}
\)
using the properties of the highest weight vector 
$|{{\bf b}^J, {\bf b}^J}\rangle$ and the Cartan matrix.
\par
%
%
Next we show that the \Kh\ forms of the \Kh\ potentials $K^J$ 
satisfy the minimal cocycle condition \eqref{Integrability}. 
For the generating $2$-cycles 
\(
C_I: \Cplx P^1 \lra G^\Cplx/ \hat H 
\) 
we take the continuous mappings
\(
v \mapsto (0,\ldots, 0, z^{\Bga_I} = v, 0, \ldots, 0).
\)
Using the properties of the projector $P^J$ and the generators 
$\Bge_{\pm i}$ we obtain that
\(
K^J|^{\:}_{C_I} (v, \bv) 
=  \gd^J_I \ln \left( 1 + \bv v \right).
\)
This implies that the integral 
\(
\int_{C_I} \go( K^J |^{\;}_{C_I} )
\)
reduces to the standard integral of the $\Cplx P^1$-\Kh\ form over
$\Cplx P^1$ itself.
As the latter is equal to $2 \gp$, we see that the minimal cocycle 
condition \eqref{Integrability} is satisfied.  
\par
%
%
Finally we determine the charges of the sections of the minimal 
line bundles, as these are the charges of the non-trivial 
matter multiplets. The charge of the $s^J$ section of the line bundle 
is the same as the charge of weight ${\bf b}^J$, we find 
\equ{
{\bf b}^J({\bf Y}^I) = (G\inv)^{JI}.
}
For the $z^{\Bga_J}$ charge of the coordinates we find
\equ{
 \Bga_J({\bf Y}^I) = \gd_J^I.
}
This shows that the minimal ${\bf Y}^J$-charge of 
a section $s^J$ of a 
line bundle over $G/H$ is $(G\inv)^{JI}$ times 
the charge of $z^{\Bga_J}$.
\\[3mm]
%
%
We now turn to the consequences of our results for 
supersymmetric model building. 
In the construction of a model we may need the transformation 
rule of a non-trivial compensating singlet to get a $U(1)$ charge 
assignment such that anomaly cancelation is ensured. 
However in general it is not guaranteed that the charge we 
need is an integer multiple of the smallest charge 
of a non-trivial singlet. 
How this restriction acts, is now illustrated by two anomaly 
free models build upon coset spaces.
\par
%
%
Ref.\ \cite{GrootNibbelink:1998tz} discusses an anomaly free
 model based on the coset $\ESO$ \cite{Achiman:1985ra}. 
The coset is parameterized by a $16$-component spinor of $SO(10)$. 
To cancel the $U(1)$-anomaly the model is extended to complete 
the $\undr{27}$ of $E_6$. 
According to the branching rule 
\equ{
\undr{27} = \undr{1}(4) + \undr{16}(1) + \undr{10}(-2),
}
this can be done by introducing a $SO(10)$ singlet with charge 
$4$ and a vector of $SO(10)$ with charge $-2$.
As discussed in ref.\ \cite{GrootNibbelink:1998tz} the vector 
$\undr{10}$ 
can be obtained from a tensor product of $2$ covariant vectors 
of $\ESO$. 
However this vector $\undr{10}$  of $SO(10)$ has charge $2$ and the 
$SO(10)$ singlet has charge $0$. 
According to the results above the minimal charge of a non-trivial 
compensating singlet is $G^{JJ} = 4/3$.
Hence we need the $3$th power of $s$ to define the singlet 
and a rescaling with the $-3$th power of $s$ of the vector 
to define the 
vector matter representation with the right charges.
Both these powers of $s$ are integers, therefore this models 
satisfies the line bundle consistency conditions.
\par
%
%
Ref.\ \cite{GrootNibbelink:1999dt} discusses another example 
of an anomaly free model that is based on the Grassmannian 
coset $SU(5)/SU(2) \times U(1) \times SU(3)$ 
\cite{Ong:1983te,Mattis:1983wn,vanHolten:1985di} with the 
chiral fermion content of the standard model.
For this model the question whether it is globally well-defined 
is more involved, as matter is not introduced as tangent vectors.
To see what the difficulty is, we give a short review of the 
introduction of matter in this model; for details we refer to 
ref.\ \cite{GrootNibbelink:1999dt}.
\par
%
%
The coordinates $Q^{ia}$ of this coset are interpreted as the 
superpartners of the left-handed quark doublet $q_L^{ia}$. 
The superpartners $L$ and $D$ of a lepton doublet $l_L^i$ and 
a (down) quark triplet $d_L^a$ were introduced by noticing that 
the transformation rule for $d Q^{ia}$ factorizes under 
infinitesimal transformations \cite{GrootNibbelink:1999dt}
\(
\gd d Q = H(Q) d Q + d Q \tilde{H}(Q)
\)
where $H(Q)$ and $\tilde{H}(Q)$ are holomorphic functions. 
This can be used to define infinitesimal transformation rules 
for $L$ and $D$:
\(
\gd L = H(Q) L
\)
and 
\(
\gd D = - \tilde{H}(Q) D
\).
Note that this transformation rule is different from the one given in 
ref.\ \cite{GrootNibbelink:1999dt}, because we $\undr{\bar 3}$ of $SU(3)$ 
in stead of the $\undr{3}$ to compile the matter representations of the 
supersymmetric standard model.
As these are only infinitesimal transformations, we do not 
obtain any information concerning global consistency.
\par 
%
%
We now show that $L$ and $D$ can be interpreted as sections 
of bundles over this coset. 
Using the transformation property \eqref{TransUxi} of $\gx(Q)$ 
with a $SU(5)$ matrix given by
\equ{
g = \pmtrx{ A & B \\ C & D },
} 
where $A \in SU(2)$, $D \in SU(3)$, $B$ a $2\times 3$ and $C$ 
a $3 \times 2$ matrix, 
we obtain the global transformation rule for $Q$:
\equ{
Q' = 
\left( A Q + B \right) \left( D + C Q \right)\inv.
\labl{GlobalTransQ}
}
By introducing the matrices $\gb = A\inv B$ and $\gg = D\inv C$ 
this can be written as
\[
\gg A\inv Q' D = \Id - 
\left( \Id - \gg \gb \right) \left( \Id + \gg Q \right)\inv.
\]
From this equation it follows that the transformation of the 
differential factorizes as 
\equ{
\frac{\der Q^{\prime\, ia} }{\der Q^{ jb} } = 
( G )^i_{\;j} 
( \tilde{G} )^{\;a}_b, 
\labl{Factor}
}
indeed 
\(
d Q' = A \left( \Id - \gb \gg \right) 
\left( \Id + Q \gg \right)\inv d Q 
\left( \Id + \gg Q \right)\inv D\inv.
\)
Now we know that the tangent bundle is well-defined, therefore 
the transition functions $G^{(ab)}$ and $\tilde{G}^{(ab)}$  
satisfy all consistency conditions by eq.\ \eqref{Factor}. 
By defining the transition functions acting on $L$ and $D$ via 
\equ{
L^{(a)} = G^{(ab)} L^{(b)} \text{ and }
D^{(a)} = (\tilde{G}^{(ba)})\inv D^{(b)},
}
well-defined bundles over $SU(5)/SU(2)\times U(1)\times SU(3)$ 
are obtained.
\par
%
%
We now turn to the line bundle constraint. 
In the normalization of the $SU(5)$ algebra employed in 
ref.\ \cite{GrootNibbelink:1999dt}, the $U_Y(1)$ charges of 
$Q$, $L$ and $D$ are resp. $5$, $3$ and $-2$. 
According to the results above the charge of a section of the 
line bundle is $G^{33} = 6/5$ times that of the charge of the coset coordinates 
$Q$, therefore the minimum charge is $6$.
The matching between the weak hypercharge $Y_w$ and the charge 
$Y$ for the quark doublets requires that $Y = 15 Y_w$. 
It can be checked that with only integer the powers of the minimal line 
bundle, it is possible to obtain the standard model hyper charges. 
We conclude that the Grassmannian standard model with 
non-linear compact $SU(5)$ symmetry can be defined 
globally.
\\[3mm]
%
%
The main objective of this work was to determine  the 
\Kh\ potential that satisfy the 
minimal quantization condition \eqref{Integrability} for a 
general \Kh ian compact coset space $G/H$. 
The BKMU-projector $P^J$  projects on 
the highest weight vector of a representation with highest 
weight ${\bf b}^J$ with the Dynkin labels 
\(
(b^J)_j = \gd^J_j.
\)
This projector was used to define the \Kh\ potential $K^J$ 
(eq.\ \eqref{KJ}) and its properties showed that this \Kh\ 
potential satisfies the minimal cocycle conditions. 
The minimal charge of a section $s^J$ of a line bundle 
associated with $\Bgt_J$ was found to be half of the charge 
of $z^{\Bga_J}$.
In this way we obtained an constructive proof a theorem by 
Borel-Weil discussed by Serre \cite{Serre:54} that classifies 
the elements of the (co)homology groups of $G/H$.
\par
The importance of these results for supersymmetric model 
building lies in the fact that they provide an additional 
restrictions on the charge assignment of supersymmetric matter 
coupling and may be in conflict with the requirement of anomaly 
cancelation. 
This has been illustrated by the review of two anomaly free 
models containing a coset space at their core. 
The $\ESO$ model satisfies the consistency condition. 
The analysis for $SU(5)/SU(2)\times U(1)\times SU(3)$ was 
more involved. 
First we showed that the scalars $L^i$ and $D^a$ are sections 
of well-defined bundles:
using the fact that the transition function of the tangent 
bundle factorizes, transition functions for these bundles were 
obtained. 
We have seen that it is possible to obtain a Grassmannian standard 
model that satisfies the line bundle constraint as well. 
\par
As is clear from these examples the consistent definition of 
a line bundle provides a stringent restriction on 
supersymmetric $\gs$-models, in particular those based on 
coset spaces. 
With this in mind, it is interesting to study what kind of other 
supersymmetric models are allowed and what their phenomenology is.
\\[3mm]
%
%
{\large\bf Acknowledgments}
\\[3mm]
I would like to thank C. Hofman for explaining the importance 
of quantization condition for line bundles to me, 
J.J. Duistermaat for introducing me to classification of 
line bundles of $G/H$ coset spaces by the work of Borel and 
Weil as is discussed in Serre's Bourbaki notes \cite{Serre:54} 
and
J.W. van Holten for valuable discussions while this work was 
in progress.
%
%
\providecommand{\href}[2]{#2}\begingroup
\endgroup

\end{document}